\documentclass[conference]{IEEEtran}
\usepackage{amsmath}
\usepackage{graphicx}
\usepackage{url}
\usepackage{mcode}
\usepackage{amsmath}
\usepackage{color}
\usepackage{xfrac}
\usepackage{flushend}
\usepackage{subfiles}
\usepackage{float}

\makeatletter
\renewcommand*\env@matrix[1][*\c@MaxMatrixCols c]{%
  \hskip -\arraycolsep
  \let\@ifnextchar\new@ifnextchar
  \array{#1}}
\makeatother

\begin{document}

\title{A Study on Quantum Radar Technology Developments and Design Consideration for its integration}
\author{
\IEEEauthorblockN{Manoj Mathews
\\}
\IEEEauthorblockA{Dept. of Electrical and Computer Engineering\\
Rowan University\\
201 Mullica Hill Rd, Glassboro, New Jersey, 08028 USA\\
mathew48@students.rowan.edu}
}
\maketitle

\IEEEpeerreviewmaketitle

\begin{abstract}
This paper presents a study on quantum radar technology developments, design Consideration for its integration, and quantum radar cross section (QRCS) based on quantum electrodynamics and interferometric considerations. Quantum radar systems supported quantum measurement can fulfill not only conventional target detection and recognition tasks but also capable of detecting and identifying the RF stealth platform and weapons systems. The development of radar technology is of the utmost importance in many avenues of research. The concept of a quantum radar has been proposed which utilizes quantum states of photons to establish information on a target at a distance. A photon, or a little cluster of photons, is distributed towards the target. The photons are absorbed and re-emitted from the target and into the receiver. The measurement process may be executed in two alternative ways. One can perform an interferometric measurement (or phase measurement) on the photon, or one can simply count the number of photons that return. the previous method is named Interferometric Quantum Radar, and therefore the latter method is termed Quantum Illumination. For either of those methods, one can use stationary quantum states of photons or use entangled states. it's been shown that entangled states provide the most effective possible boost in resolution, achieving within the ideal case. The benefit of using quantum states is that they exhibit extra degrees of correlation by which to get information compared to classical methods. These extra correlations (called quantum correlations) serve to boost the resolution and signal/noise (SNR) that may be achieved within the radar system. To obtain information, a relation study is done between the two photons. In order to beat the diffraction limit, coherent state quantum radar depends on the use of coherent state photons and a quantum detection scheme. This paper also reviews the principles of design and operation of quantum applied trends.
\end{abstract}


\section{Introduction}
Quantum radar could be a new exemplar that exploits quantum phenomena to reinforce the resolution of a radar system during which it becomes more sensitive than its classical counterpart. There are two emerging methods to accomplish this, and every method has its own unique characteristics. Quantum phenomena, such as quantum entanglement, as a kind of natural resource, have been widely used in quantum computation and quantum communication. It provides a solid material basis for the advancement of the science of quantum knowledge. Quantum metrology makes use of quantum phenomena to improve measurement sensitivity. Theoretical research shows that quantum measurements can break through the normal quantum limits and calculate supersensitivity. Quantum information technology will become the key to improving the efficiency of sensor systems in the near future. Radar systems Based on quantum measurement are not only able to perform conventional target detection and recognition but also able to accomplish the detection and identification of RF stealth platform and weapons systems. Due to the clear connection between entangled photons, any attempt to deceive quantum radar would be exposed. On the other hand, stealth aircraft can be tracked at great distances due to the supersensitivity of quantum measurement, and even for stealth aircraft such as F-22 and B-2, the detection range could be reached at several hundred to several thousand kilometers. Quantum radars will lead a new technology revolution in electronic warfare just like the RF stealth technology did in the past 20 years. In addition to military applications, quantum measurement technology can be widely used in interplanetary defense and space exploration.

\section{Quantum Radar Theory}
 \subsection{The Fundamental Limits of Quantum Radar}
Quantum radar can be defined as a type of stand-off measuring device that uses microwave photons, optical photons, and quantum phenomena to enhance the efficiency of target detection and recognition. The biggest advantage of quantum radars over traditional classical radars is inherited from the nature of entanglement states used in the transmitting signals. Examples of this kind are interferometric quantum radars and radars with quantum illumination. The so-called quantum entanglement refers to the strong correlations between quantum systems which are non-classic and non-local. Theoretically, no matter how far the gap divides, including one on the planet and the other on the edge of the Milky Way, the peculiar bond between two intertwined states exists. For example, when one of them is manipulated, calculated, and the other changes to the corresponding state immediately. The correlation between the entangled states can’t be explained classically and therefore Einstein called it a kind of ‘Spooky Action at a Distance’. In any event, most detection strategies are based on the exact same principle: perform collective measurements after highly correlated states have been injected into the system. Such is the case with the interferometric quantum radar that crosses the Heisenberg limit by using strongly entangled states. Many improved imaging developments have been made possible by recent innovations in quantum mechanics. Entangled photon-number (N00N) states have allowed Heisenberg-limited phase measurement and led to the development of radar systems with quantum-enhanced resolution. The fundamental limit given by Heisenberg’s principle based on N00N states phase measurement is as below.
\begin{equation} \label{eq1}
\begin{split}
\Delta\varphi  & \geq \frac{1}{(N)} \\
\end{split}
\end{equation}
Where $\Delta\varphi$ represent the phase fluctuation, N is the entangled photons in the quantum system,  means taking the average. The Heisenberg’s limit does not depend on measurement strategies and it is unavoidable. At the same time the sensitivity of most modern sensors is bounded by the standard quantum limit (ie. Shot noise limit). In the case of standard quantum limit the phase measurement resolution is as below.
\begin{equation} \label{eq1}
\begin{split}
\Delta\varphi  & \geq \frac{1}{\surd(N)} \\
\end{split}
\end{equation}
The Super-sensitivity Regime is called a regimen of variables for which the sensitivity of a sensor exceeds the value imposed by the normal quantum limit, and the sensor is said to be Supersensitive. It should be remembered that the metrology of the Heisenberg limit is still very difficult to achieve in a functional framework. There is a wide range of variables at these high accuracy scales that have major contributions and should not be ignored, such as thermal noise, platform vibrations,  imperfect alignment of optical elements, etc. What is more exciting about quantum radar is that every third party except for the radar transmitter and receiver will not accurately copy or secretly change the quantum states due to the Heisenberg Uncertainty Principle and the strong global association between entangled states. Also, the most sophisticated stealth aircraft such as B-2 and F-22 have nowhere to hide under the surveillance of quantum radar systems, making any existing jamming and deception approach invalidation to quantum radars.

 \subsection{Standoff Quantum Sensor Classification }
Standoff quantum sensing architectures can be classified according to the type of quantum phenomena exploited by the system. The three basic categories are the following:\\
Type 1: The quantum sensor transmits un-entangled quantum states of light.\\
Type 2: The quantum sensor transmits classical states of light, but uses quantum photo-sensors to boost its performance.\\
Type 3: The quantum sensor transmits quantum signal states of light that are entangled with quantum ancilla states of light kept at the transmitter.\\
The principal examples of quantum radar systems are the single photon quantum radar and the entangled photon quantum radar. These are Type-1 and Type-3 sensors, respectively. On the other hand, quantum LADAR is an example of a Type-2 sensor.
\subsection{Single-Photon Quantum Radar}
\begin{figure}[H]
\centering
\includegraphics[width=1\linewidth]{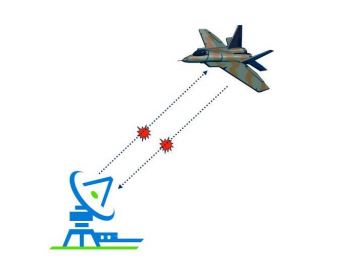}
\caption{ Single-photon radar :A single photon is emitted towards a target, and subsequently the photon is reflected back to the receiver. }
\label{Figure1a1b}
\end{figure}
Quantum radar can be divided into two specific groups according to the signals emitted by transmitters:\\
i. Quantum radars use un-entangled signals. \\
ii. Quantum radars use entangled photons.\\ 
Single-photon quantum radars(in Fig. 1) are Type-1 sensors. These systems work in a manner close to that of a classical radar. Its transmitter generates signal pulses contain a single photon (by single it means in an average way) and sends pulses towards a target.And its receiver attempts to collect the reflected photons and detects photon-counting. Quantum radar is not a true quantum radar that utilizes the quantum phenomenon purely with a single photon. But the unintended benefit of single-photon quantum radar is that if low photon number pulses are used for target detection, the radar cross-section is larger.That is to say, the target appears to look bigger when using single-photon pulses than use classical light beams.
\subsection{Entangled-Photon Quantum Radar}
Through the use of entangled states of light, the greatest advantage of quantum radars is obtained.These are Type-3 sensors. As shown in Figure 2, an entangled pair of photons is generated.One photon is sent to the target and the other is retained in the radar device. The outgoing photon is reflected by the target and received by the radar subsequently. At this stage to improve detection efficiency, the correlations embedded in the entangled states are exploited. The Interferometric Quantum Radar and Quantum Illumination are examples of such devices.
\begin{figure}[H]
\centering
\includegraphics[width=1\linewidth]{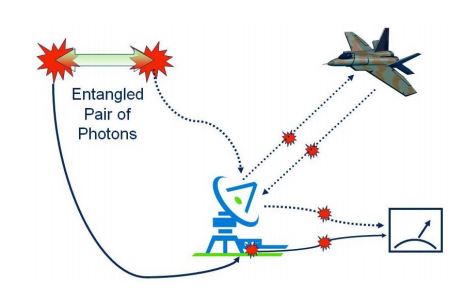}
\caption{Quantum radars based on entangled photons:An entangled pair is produced, one of the photons is sent towards the target while the other is kept in the system. The correlation between these photons can be exploited to increase the performance of the device.}
\label{Figure1a1b}
\end{figure}
\subsection{Interferometric Quantum Radar}
In an interferometric quantum radar, as shown in Fig. 2, an entangled pair of photons is generated. One of these is sent towards the target and the other one is held inside of the radar receiver. Very quickly, the photon sent out is reflected back and is received by the radar. Then the correlations embedded in the entangled states are exploited to increase detection performance.Type-2 quantum radars are interferometric quantum radars and quantum illumination.
\begin{figure}[H]
\centering
\includegraphics[width=1\linewidth]{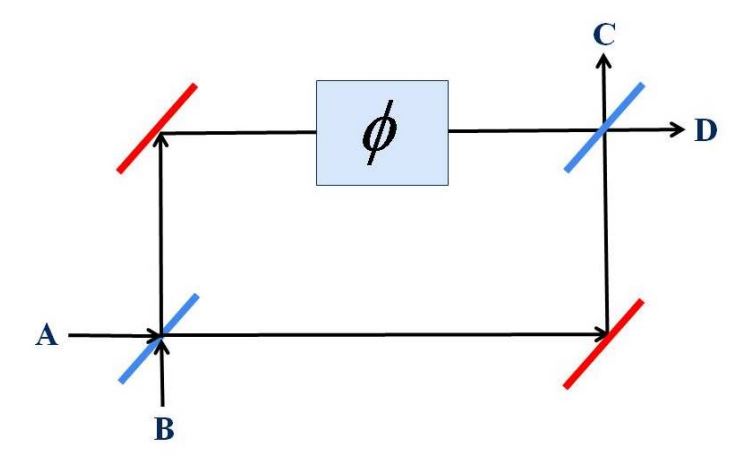}
\caption{Mach-Zender interferometer with input ports A and B, output ports C and D,and a phase delay $\phi$ in one of the arms}
\label{Figure1a1b}
\end{figure}
An interferometric quantum radar makes phase measurements by interferometers like Mach-Zender interferometers (see Fig.3). The input light field in a Mach-Zehnder interferometer is separated by a beam splitter into two distinct paths and recombined by another beam splitter. The phase divergence between the two paths containing the target distance information is then calculated by balanced detection of the two modes of performance.
\begin{figure}[H]
\centering
\includegraphics[width=1\linewidth]{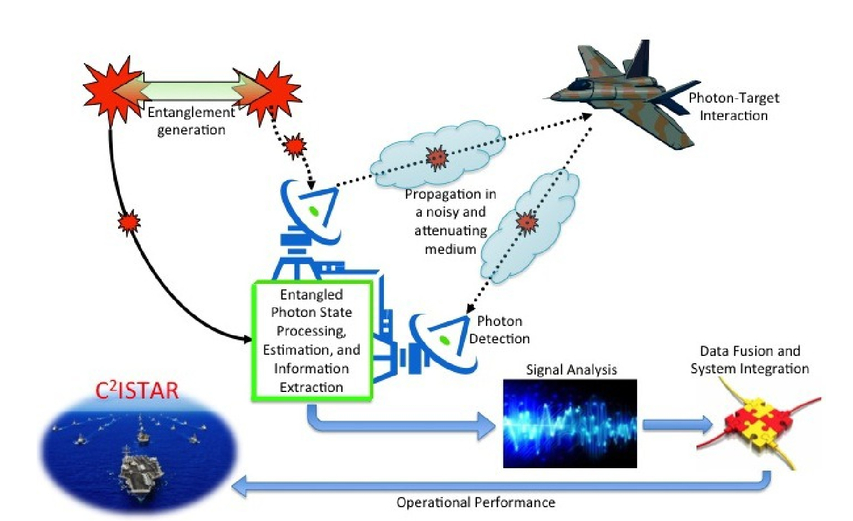}
\caption{Conceptual description of a quantum radar system.}
\label{Figure1a1b}
\end{figure}
Figure 4. shows an entangled pair of photons is produced; one photon is kept within the sensor while the other is emitted towards a region of space. In the system, an entangled pair of photons are created. The signal and the idler (or ancilla) photons are known to them. Inside the device, the idler photon is held while the signal photon is sent to a potential target via a medium. With a certain probability, the signal photon may, or may not encounter the target. If the target is not encountered by the signal photon, it will continue to spread in space. All measurements performed by the detector would be of noise photons in such a situation. On the other hand, if the target is present and the signal photon is “bounced back” towards the detector, then it will be detected with a certain probability. In a context, because of the quantum correlations due to the entanglement, the signal photon is "tagged" and it would thus be "easier" to correctly classify it as a signal photon rather than misidentify it as a noise photon.
\subsection{Quantum Radar Based on Quantum Illumination}
The setup of the Quantum Illumination Radar is similar to the one discussed above for interferometric quantum radar, but it has a different detection strategy. One does not conduct phase measurements in the case of quantum illumination, but simple photon detection counts are adequate. Assume that a quantum illumination device is used for illuminating a target. The goal is to detect the presence of the target even in a noisy and lossy environment. And again, entanglement increases the sensitivity, but in a different way, of the detection system. Seth Lloyd at MIT first discovered quantum illumination in 2008. It is a revolutionary photonic stand-off quantum sensing technology that enhances the sensitivity of detection in noisy and lossy environments. Target illumination using entangled light can provide substantial enhancements over un-entangled light for detecting and imaging in the presence of a high level of noise and loss. The definition of quantum illumination may be clarified by a photo-taking analogy. Imagine a camera with a flashlight able to emit entangled light to take photos, and only part of these entangled photons are emitted to illuminate targets. In order to filter the noise photons that would lack correlated twins and thus significantly improve the sensitivity of imaging, the sensor compares the reflected photons with those kept within the camera. This will make it possible for radars to recognize the subtle evidence of objects usually obscured by noise. Research shows that quantum lighting with m bits of entanglements will in theory increase the efficient signal-to-noise ratio by a factor of 2 m for photon detection, an exponential improvement over un-entangled lighting. The progress continues even when the noise is so great that the detector does not withstand any entanglement. That’s another unexpected advantage of quantum illumination. Quite interestingly, only in a noisy and lossy setting is the enhancement supplied by quantum illumination using entangled photons observed. Quantum illumination can be used for ranging and imaging purposes and is not limited to any specific frequency.
\begin{figure}[H]
\centering
\includegraphics[width=1\linewidth]{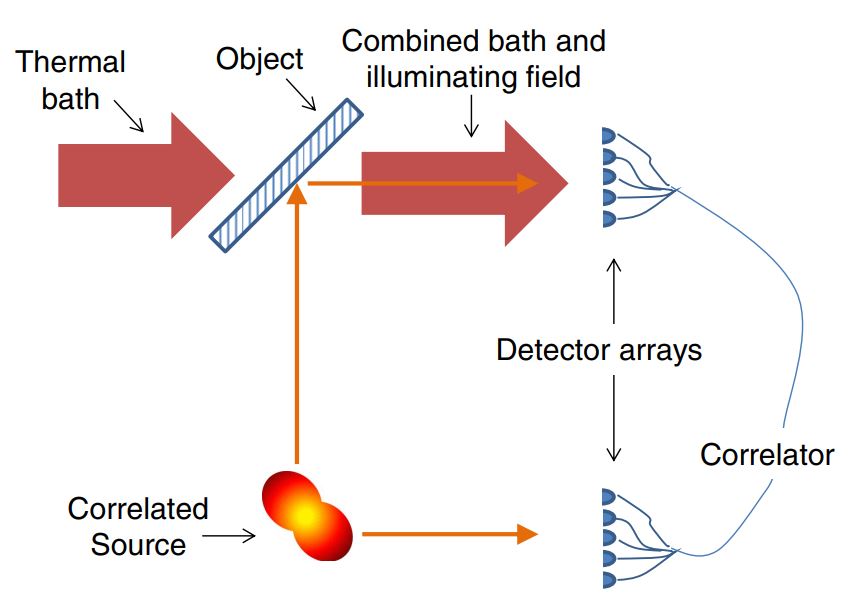}
\caption{Schematic diagram of a quantum illumination scheme.}
\label{Figure1a1b}
\end{figure}
Here a beam splitter is an element under consideration. The beam splitter will reflect a portion of the light emitted by the associated source, and the task is to detect this under the dominant thermal noise, thus discriminating against the presence of the target.

\section{Quantum Radar Physical Realization}
\subsection{Entanglement Generators}
Generation and detection of entangled photons are the key elements to realize quantum radar. Currently, standard parametric down-conversion (SPDC) is the most widely known technique to generate entangled photons in the visible frequency. The recent entanglement generation technique uses a non-linear crystal (Barium Boron Oxide-BBO) that splits an incoming photon into two entangled photons both in a lower frequency. One of the outcoming photon called signal photon is used for transmitting and the other outcoming photon called idle photon is kept locally and used for detection. Usually, the outcoming photons are generated as polarization-entangled. None of the polarization of photons is calculated before a measurement takes place. If one photon is measured as horizontal polarization and the quantum state of the other, without further measurement, one automatically switches to vertical polarization. But the SPDC method does not work to produce microwave photons that have been commonly used in classical radar, missile guidance, navigation, environment monitoring, ground monitoring, and airport traffic control. Entangled photons could be produced at a visible frequency by the use of semiconductor nanostructures. A related intra-band transfer of conduction band electrons could also produce entangled photons in the X-band in quantum dots. Photon-assisted tunneling experiments have shown that these transitions coupled to microwave photons. As a result, from spontaneous downward transitions between single-particle levels in a quantum dot, microwave photons are produced.
\subsection{Photon Detectors}
In the field of photon detection, there is a similar disparity between the visible and X-band frequency regimes. Both interferometric measurement and photon counting in the visible and near-visible regimes are well developed during the research work of quantum key distribution (QKD), quantum communications, and quantum computing. But none of the mentioned has involved the operation in X-band. Therefore there are lots of theoretical and experimental challenges in the design and development of single. photon detectors in the microwave regime. To detect single photons in the microwave regime, a novel sensing technique has recently been suggested. This detector resembles photographic film in the sense that once a photon has been absorbed by the meta-material, the device is changed into a stable and mesoscopic distinguishable state. Its limited operating bandwidth is one possible issue with the proposed design. Another potential concern is that there will be incoming photons that will not be counted if the decay process takes place. However, recent experiments appear to suggest that this process happens at a rate of a few MHz, and therefore it will affect long-wave packets. Whether this problem may pose a significant limitation to the use of these instruments for quantum radar applications in the X-band remains to be seen.

\section{Scheme For Entangled Photon Quantum Radar}
This section discusses a design scheme proposed by Bassyouni for an entangled quantum radar. There have been a few different schemes suggested, but they are all quite similar. The block diagram for the entanglement radar scheme is shown in Figure 6. The suggested method for entangled photon production is that of spontaneous parametric down-conversion (SPDC). A signal photon and an idler photon are produced by the SPDC process, which is both entangled. The signal photon beam is sent towards the target via an antenna. The idler photon beam, which is used to detect the quantum state, is sent directly into the idler detector array. The idler detector output appears simultaneously and waits in the memory until the returned signal is processed in the receiver and detected in the signal detector array as soon as the transmitted signal reaches the target. To process the target parameters, the signal and idler detector outputs are added to the coincidence estimator and signal processing modules. To calculate more than one parameter for a given state, the signal is split into multiple signals through a beam splitter array. Each signal is then sent into a detector that is designed to detect certain properties of the signal. This is shown in Figure 4. Only one quantum state can pass through one of the many beam splitting paths. But on the theory of sending out multiple entangled photons, the entanglement radar works. They will randomly go through various paths of the beam splitter array upon receipt of these photons, and finally, all paths will be taken.It is necessary to note that because the SPDC process emits optical photons, this particular design is confined to the optical regime. However, if the SPDC process is replaced with a process of microwave photon generation, the same design may be implemented.
\begin{figure}[H]
\centering
\includegraphics[width=1\linewidth]{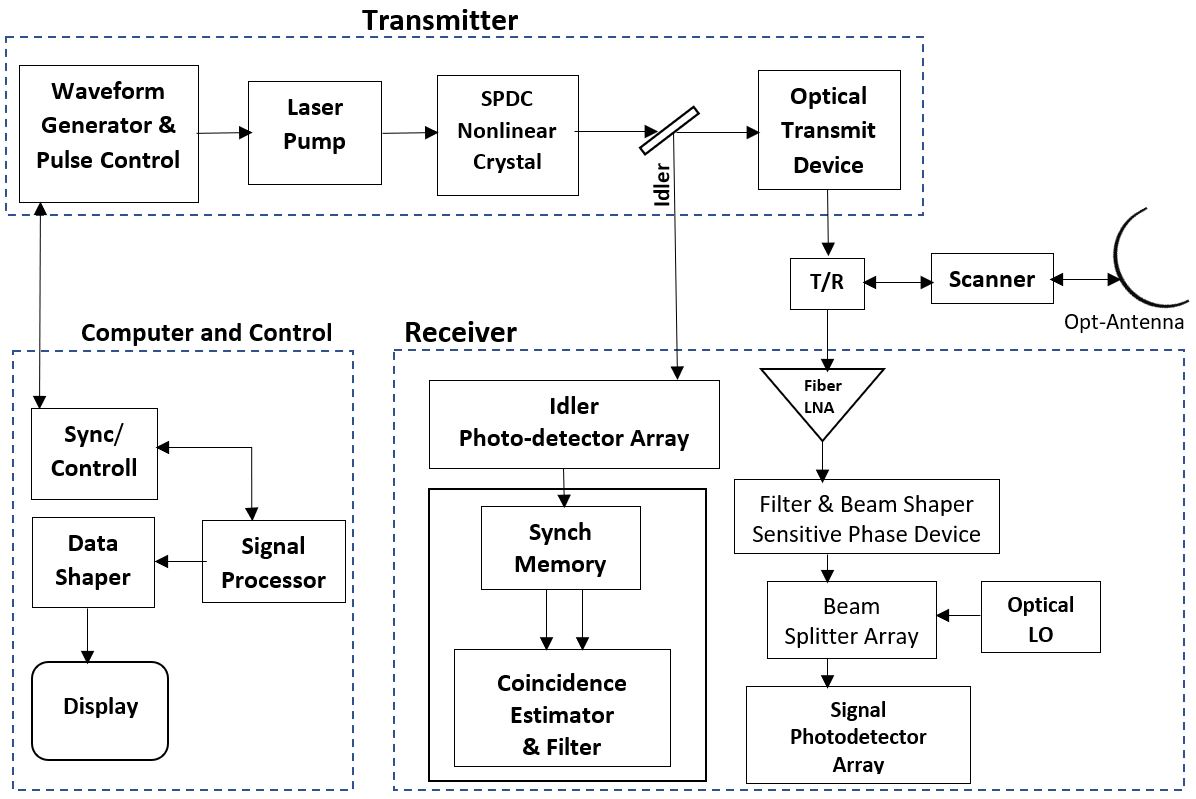}
\caption{Block diagram of an entanglement Photon quantum radar}
\label{Figure1a1b}
\end{figure}
\begin{figure}[H]
\centering
\includegraphics[width=1\linewidth]{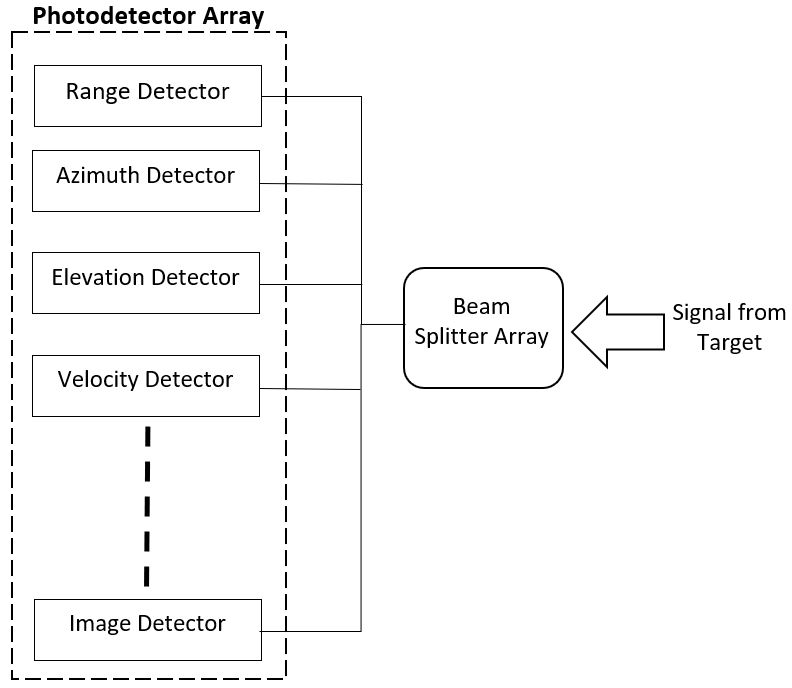}
\caption{Array detector for obtaining information about the targets parameters}
\label{Figure1a1b}
\end{figure}
A spontaneous parametric down-conversion of the photon source is (SPDC).The signal photon passes out of the antenna and is sent into the receiver by the idler photon, where it is retained until the return signal arrives. Then the two photons are sent into the coincidence estimator and data is collected from the data received.\\
Array detectors are for the purpose of obtaining target parameter information. Through a beam splitter, the echo signal is split into multiple separate signals.From here to ascertain data, each split signal will go into a different measurement unit.

\section{Development Of Quantum Radar}
The concept of using the method of quantum detection to enhance target detection sensitivity appeared very early on. As early as 1991, the U.S. navy proposed a quantum detector patent exploit to increase the sensitivity of conventional radars.This patent proposed the use of tunable Rydberg detectors to improve the detection sensitivity to the quantum level of classic radar systems. The patent says that the transmitter sends classical pulses, but from today's point of view, it is not a proper quantum radar device. In 1991, the E91 quantum key distribution protocol introduced the use of quantum entangled states as signal sources. Then as a sort of mystical natural resource, quantum entanglement started to attract the broad attention of scientists. The extensive study. For the last 20 years, systematic research and the use of quantum entanglements have contributed to a fundamental shift in information science. A few years later, quantum entanglement was proposed to be used to break through the normal measurement quantum limits and hit the degree of supersensitivity measurement. And quantum entanglement has also been used to improve imaging resolution. In 2005, Lockheed Martin Corporation suggested a quantum radar device based on a multi-particle source of quantum entanglement. The radar signal is composed of several entangled particles with varying frequencies in this scheme. And the radar transceiver is capable of somehow manipulating any frequency. Relatively short wavelengths of entangled particles strive to achieve high-resolution imaging, whereas longer wavelengths of entangled particles aim to achieve long-range target detection. The key benefit of this quantum radar theory is that it resolves the contradiction between resolution and detection range in conventional radars and makes radar systems capable of seeing clearly and far at the same time. Another type of quantum radar scheme based on quantum illumination was developed shortly after the grant of this U.S. patent in 2008 and submitted to the United States Patent Office in 2009. In 2012, this quantum illumination radar patent was issued. \par
To date, not only does quantum radar remain a theoretical proposition, but it has also been adopted in laboratories for preliminary implementation. The first quantum radar demonstration in the laboratory that confirmed the anti-stealth capability of quantum radar was performed by Boyd and his research group. They constructed an imaging device that uses the location or time-of-fight information of a photon to image an object while using the polarization of the photon for protection. This skill helps the radar to produce an image that is safe from an attack in which the imaged object intercepts and resends modified information to the imaging photons. The delicate quantum state of the imaging photons must be disrupted by the target, thereby producing statistical errors that expose its behavior. In 2008, Lloyd suggested the theoretical possibility of quantum illumination imaging in a high-level noise context, but it is difficult to evaluate because of the fragility of entanglement. It is hard to test the idea. Lloyd's proposal was experimentally demonstrated by Marco Genovese, a physicist at the National Institute for Metrological Research in Turin, Italy, and his colleagues, based on photon-number correlations in the laboratory. This will make it possible for sensors to distinguish the subtle evidence of objects currently hidden by noise. There are also other quantum radar systems in the theory stage and waiting for further experimental verifications.

\section{Quantum Radar Cross Section}
  In reality, targets will present complicated geometries that will represent a complex pattern of incoming photons. The radar cross section \(\sigma _C\) is used within the realm of classical radar theory to assess the "radar visibility" of a particular target.As quantum radars emit a handful of photons at that time, photon-atom scattering procedures controlled by the laws of quantum electrodynamics define the radar-target interaction in this regime.As such, using the same \(\sigma _C\) to describe the visibility of a target illuminated by a quantum radar is theoretically inconsistent.As a result, to objectively calculate the "quantum radar visibility" of a specified target, the concept of a quantum radar cross section \(\sigma _Q\) needs to be established.That is, in the scenario where the targets are not perfectly reflective objects and the radar signal is made of a handful of photons, we need to describe \(\sigma _Q\)  to evaluate the performance of quantum radars.
 
 \subsection{Desired properties of quantum radar cross section (\(\sigma_Q\))}
The basic properties that are desired in a conceptually robust definition of \(\sigma_Q\). In simple analogy to the classical radar cross section, the quantum radar cross section should have these properties:\\
\textbf{ Operational Meaning:} The reason to define \(\sigma_Q\) in the first place is to have an objective measure of the quantum radar visibility of a target.\\
\textbf{Energy Conservation:} As with \(\sigma_C\), \(\sigma_Q\) should entail energy conservation in the optical regime when absorption effects are ignored.\\
\textbf{Strong Dependencies:} Similarly, to \(\sigma_C\), it is desirable that \(\sigma_Q\) strongly depends on properties of the target: geometry (absolute and relative size, shape and orientation), as well as its composition (material properties).\\
\textbf{Weak Dependencies:} In the same manner, it is desired that \(\sigma_Q\) is approximately independent on the properties of the radar system. That is, \(\sigma_Q\) should depend very weakly on the strength, architecture, physical implementation, and range of the radar system.\\
\textbf{Multi platform Comparison:} To better understand the advantages and disadvantages of quantum radars, it would be desirable to be able to directly compare \(\sigma_C\) and \(\sigma_Q\).\\
\textbf{Asymptotic Behavior:} In the large photon limit, the quantum realm gives way to classical physics and \(\sigma_Q\) should be proportional to \(\sigma_C\).
\begin{equation} \label{eq1}
 \lim_{n_\gamma \to\infty} \sigma_Q \propto \sigma_C
\end{equation} 
\subsection{Quantum Radar Equation}
It is reasonable to define \(\sigma_Q\) in analogy to \(\sigma_C\) as:
\begin{equation} \label{eq1}
\sigma _Q \equiv \lim_{R\to\infty} 4\pi R^2 \frac{(I_s(x_s,x_d,t))}{(I_i(x_s,t))}
\end{equation} 
And if we assume energy conservation in the optical regime, it is possible to approximate \(\sigma_Q\) for a mono static quantum radar with:
\begin{equation} \label{eq1}
\sigma _Q  \approx4\pi A\bot(\theta,\phi) \lim_{R\to\infty} \frac{(I_s(x_s = x_d))}{ \int_{0}^{2\pi}\int_{0}^{\pi} (I_i(x_s,x_d)) sin\theta_d d\theta_d d\phi_d}
\end{equation} 
Where the expectation value at the receiver of the scattered intensity is taken. Unfortunately, for the analytic analysis of quantum radar, this "simplified" expression for \(\sigma_Q\) remains problematic. In general, to elucidate the behavior of the quantum radar cross section, numerical methods have to be used.\\
Based on the definition of quantum radar cross section,
\begin{equation} \label{eq1}
\sigma _Q = \lim_{R\to\infty} 4\pi R^2 \frac{(I_s)}{(I_i)}
\end{equation} 
Using the expressions for the intensities we get in the large R limit. 
\begin{equation} \label{eq1}
\begin{split}
 (I_s) & \approx \frac{(I_i)\sigma_Q}{4\pi R^2 }
\end{split}
\end{equation} 

\begin{equation} \label{eq1}
\begin{split}
  & \approx \frac{\epsilon _0 ^2\sigma_Q}{4\pi R^4 }
 \end{split}
\end{equation} 

\begin{equation} \label{eq1}
\begin{split}
(I_s) & \approx \frac{4\pi \epsilon _0 ^2\sigma_Q}{(4\pi)^2 R^4 }
  \end{split}
\end{equation}
The above expression resembles the classical radar equation making replacements:
\begin{equation} \label{eq1}
\begin{split}
 P_t ^Q =  4\pi \epsilon _0 ^2
 \end{split}
\end{equation} 
\begin{equation} \label{eq1}
\begin{split}
 P_r ^Q =  (I_s) A_r
 \end{split}
\end{equation} 
where $ P_t^Q $is the transmitted power of the quantum radar and $P_r^Q$ is the received power at the quantum radar. Then the Quantum Radar Equation is:
\begin{equation} \label{eq1}
\begin{split}
 P_r ^Q =  \frac{P_t ^Q A_r\sigma_Q } {(4\pi)^2 R^4 }
\end{split}
\end{equation}

\section{Analysis of Quantum Radar Cross Section}
 \subsection{\(\sigma _Q\) for Rectangular Targets}
Referring to the recent researches Fig. 8 shows a flat rectangular plate of size A=axb, to be in the XY plane, this means that to observe it at the principal angles. A\(\bot\) (projected cross sectional area of the target which is a function of \(\theta_i\) and \(\phi_i\)), is changing based on the viewing angle for each object of interest. For flat objects, the projected cross sectional area should be zero at the extreme angles (looking from a side view i.e., when \(\theta\) = \(\pi\)/2 and attains a maximum when looking at normal incidence (i.e., when \(\sigma _C\)= 0 or \(\pi\)). This can be modeled by the following expression:
\begin{equation} \label{eq1}
A(\theta) = A \perp|\cos \theta|
\end{equation}
where A\(\bot\) is the cross sectional area of the object at normal incidence, and \(\theta\) varies between -\(\pi\)/2) and \(\pi\)/2).The absolute value ensures that the projected area is always positive.
 \(\sigma _C\) for Rectangular Targets,
\(F(V(x'))\)represents the Fourier transform of \(V(x')\)
\begin{equation} \label{eq1}
F(V(x')) = \iint (V(x')e^{iKx'}dS 
\end{equation}

\begin{equation} \label{eq1}
\sigma _Q = \frac{4\pi A\bot(\theta,\phi)|F(V(x'))|^{2}} {\int_{0}^{2\pi}\int_{0}^{\pi} |F(V(x'))|^{2} sin\theta' d\theta' d\phi'}
\end{equation} 

\begin{figure}[H]
\centering
\includegraphics[width=1\linewidth]{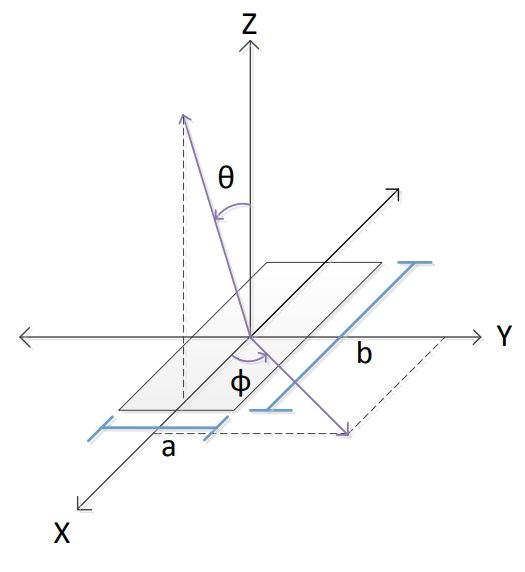}
\caption{Geometry of the rectangular plate.
.}
\label{Figure1a1b}
\end{figure}
For a 2D rectangular plate, of width a and height b, Equation (14) becomes the following:

\begin{equation} \label{eq1}
F(V(x')) = (ab)\frac{\sin{K_x a}} {K_xa}
\end{equation} 
Referring to the recent researches,we can summaries that the value of \(\chi(k,a,b)\) is only valid at high frequencies. Therefore  for high frequencies in comparison to the target size, the QRCS equation becomes.
\begin{equation} \label{eq1}
\sigma _Q = \frac{4\pi(ab)^2}{\lambda^2}|\cos\theta|\Bigg(\frac{\sin(ka\sin\theta)}{kasin\theta}\Bigg)^2
\end{equation} 

\begin{figure}[H]
\centering
\includegraphics[width=1\linewidth]{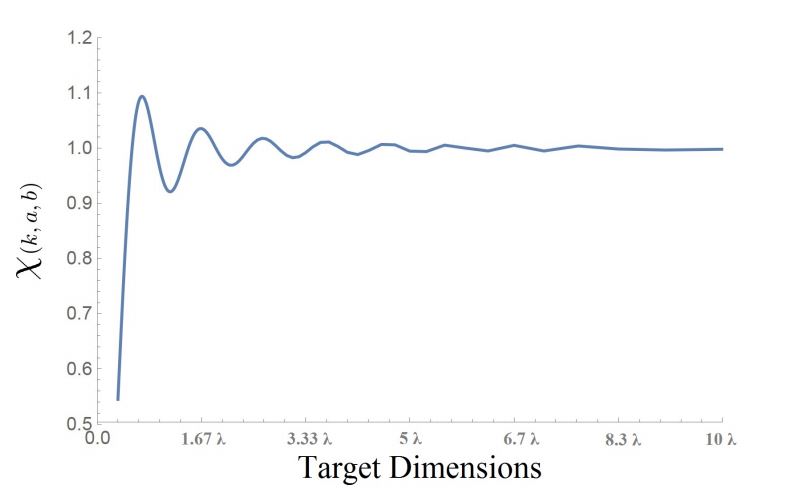}
\caption{\(\chi(k, a, b)\) Vs Target Dimensions}
\label{Figure1a1b}
\end{figure}
Fig.9 shows the Value of \(\chi(k, a, b)\) for a single frequency of a square plate over several values of dimensions. In the high frequency regime (in relation to object size), the value of \(\chi(k, a, b)\) is always approximately equal to 1. As one gets to lower frequencies (in relation to object size), the value of \(\chi(k, a, b)\) oscillates around one, and eventually drops to far less than 1. This behavior is due to the underlying high frequency approximation breaking down.\\
When we look into the case when all of the targets electric dipoles are aligned, the value of \(\chi(k, a, b)\)  will not be equal to 1. When we directly compare the equations of the CRCS and the QRCS and determine why the QRCS provides a side lobe enhancement over the CRCS.
The CRCS expression is given by,
\begin{equation} \label{eq1}
\sigma _C = \frac{4\pi(ab)^2}{\lambda^2}|\cos^2\theta|\Bigg(\frac{\sin(ka\sin\theta)}{kasin\theta}\Bigg)^2
\end{equation} 

The plots of these two equations for a plate size of \(4\lambda\)×4\(\lambda\) is shown in Figure 10. We see that the two equations are very similar. From Equations (17) and (18), we note that the QRCS equation contains a \(|\cos \theta|\)term, while the CRCS equation contains a \(\cos^2\theta\) term. This term is the origin of the QRCS sidelobe advantage can be observed by comparing plots of the
two cosine functions in Figure 11.
\begin{figure}[H]
\centering
\includegraphics[width=1\linewidth]{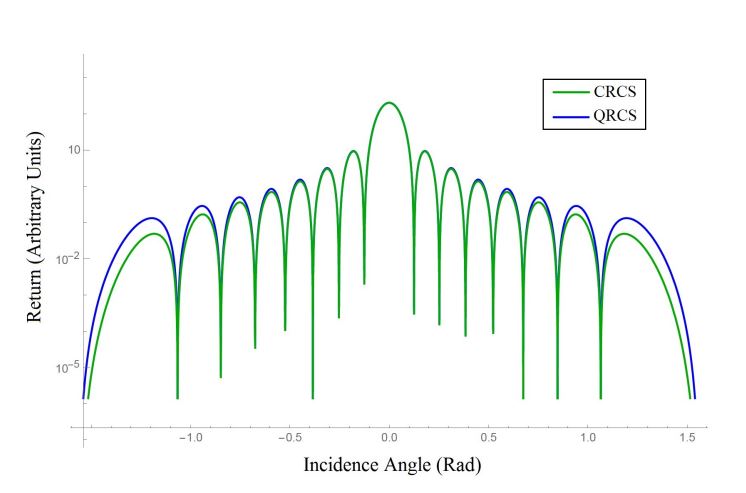}
\caption{Comparison of the QRCS and CRCS equations. The QRCS has a larger response at large scattering angles. Plate size is 4\(\lambda)\) × 4\(\lambda)\).}
\label{Figure1a1b}
\end{figure}

\begin{figure}[H]
\centering
\includegraphics[width=1\linewidth]{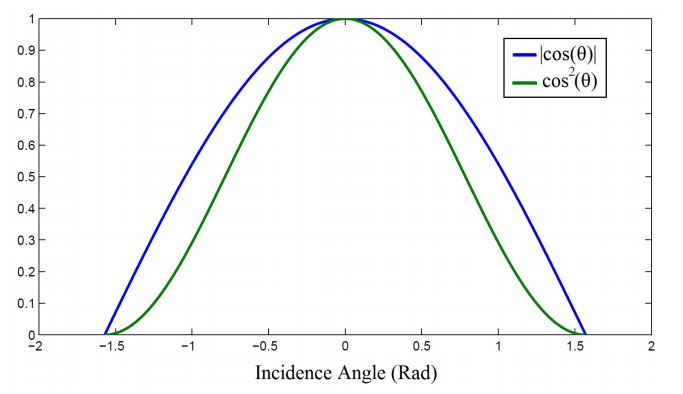}
\caption{Comparison of \(|\cos \theta|\) and \(|\cos^2 \theta|\). The plot of \(|\cos \theta|\) has a larger response over the regions that correspond to the sidelobes in the RCS expressions. This is the origin of the sidelobe advantage in the QRCS.}
\label{Figure1a1b}
\end{figure}

In the QRCS, it is obvious to see that the term \(|\cos \theta|\) emerges from the equation for the projected cross sectional area,\(A\bot\). This projected area term originates from integrating the incident expected intensity over the surface of the target during the derivation of the QRCS equation. The QRCS manifests as a result of quantum interference from the atoms on the surface of the target which (like the infinitesimal currents on the target) scatter isotropically. The difference here is that the atoms that make up the target are not vector quantities and do not need to be decomposed into components for the integration. The result is one single surface integral instead of multiple surface integrals over the different component interactions. The difference between the cosine and cosine squared term can be explained in a more physical manner, which provides additional insights. Physically, the infinitesimal currents induced on the target during classical scattering act as small antennas, which then radiate and sum together in a particular direction. In the
quantum case, the target response is from many isolated atomic transition events, which produce wave functions for a photon, which then sum together in a particular direction.\\

\section{Conclusion}
Quantum radar is a promising technology that will have a powerful effect on civilian and military environments. Although quantum sensing technology isn't as mature as quantum cryptography or quantum communications, it's not as challenging as quantum computations. In General quantum radars have already got their basic feasibility within the aspects of theory and realization, and there seems no insolvable scientific obstacle against it for the instant. In contrast to its classical radar equivalent, Quantum radar offers the simplest way to significantly increase the resolution. The power to trace aircraft and weapons is additionally included within the case of entanglement radar. Due to many problems, entanglement radar is sort of a bit out of control. One major issue is that the undeniable fact that within the microwave regime, a reliable and stable single entangled photon transmitter has yet to be produced. All previously proposed models had a photon source coming from random methods of parametric down-conversion. Research works shows that the larger the value of QRCS is, the more powerful is the scattering capacity for the incident photons; accordingly, a better detection performance can be obtained between quantum radar and targets. Also by solving
more problems on target scattering character, such as how to
obtain more information about the target by identifying the
states of photons

\end{document}